# THE CHALLENGE OF EMPLOYEE MOTIVATION IN BUSINESS MANAGEMENT

Anna KASPERCZUK[1*], Michał ĆWIĄKAŁA[2], Ernest GÓRKA[3], Dariusz BARAN[4], Piotr RĘCZAJSKI[5], Piotr MRZYGŁÓD[6], Maciej FRASUNKIEWICZ[7], Agnieszka DARDZIŃSKA-GŁĘBOCKA[8], Jan PIWNIK[9]

[1] Bialystok University of Technology, Faculty of Mechanical Engineering; a.kasperczuk@pb.edu.pl, ORCID: 0000-0002-5919-5346
[2] I'M Brand Institute sp. z o.o.; m.cwiakala@imbrandinstitute.pl, ORCID: 0000-0001-9706-864X
[3] WSB - National-Louis University, College of Social and Computer Sciences; ewgorka@wsb-nlu.edu.pl, ORCID: 0009-0006-3293-5670
[4] WSB - National-Louis University, College of Social and Computer Sciences; dkbaran@wsb-nlu.edu.pl, ORCID: 0009-0006-8697-5459
[5] MAMASTUDIO Pawlik, Ręczajski, spółka jawna; piotr@mamastudio.pl, ORCID: 0009-0000-4745-5940
[6] Piotr Mrzygłód Sprzedaż-Marketing-Consulting; piotr@marketing-sprzedaz.pl, ORCID: 0009-0006-5269-0359
[7] F3-TFS sp. z o.o.; m.frasunkiewicz@imbrandinstitute.pl, ORCID: 0009-0006-6079-4924
[8] Bialystok University of Technology, Faculty of Mechanical Engineering; a.dardzinska@pb.edu.pl, ORCID: 0000-0002-2811-0274
[9] WSB Merito University in Gdańsk, Faculty of Computer Science and New Technologies; jpiwnik@wsb.gda.pl, ORCID: 0000-0001-9436-7142
* Correspondence author

**Purpose:** This paper examines the role of employee motivation in effective business management. It also explores the impact of financial and non-financial motivators on employee engagement.
**Design/methodology/approach**: The research used a quantitative methodology and an online survey of 102 individuals. Statistical analyses, including variance analysis and correlation analysis, were conducted to identify significant patterns and differences in motivation levels.
**Findings:** Financial motivators, particularly bonuses for achieving targets, were identified as the most effective. Non-financial motivators, such as flexible work schedules and additional days off, also showed high effectiveness in enhancing motivation. Significant differences in motivation levels were observed by gender, age, and length of service.
**Research limitations/implications**: The study is limited to a specific demographic and geographic scope. Future research could explore diverse cultural and occupational contexts.
**Practical implications:** Combining financial and non-financial motivators can effectively increase employee satisfaction and engagement.
**Social implications:** Fostering effective motivation practices contributes to stronger family-company relationships. Companies adopting such strategies set benchmarks for the best workplace.
**Originality/value:** This study provides insights into the balance of financial and non-financial motivators in shaping employee motivation. It offers actionable recommendations for HR managers and organisational leaders.

**Category of paper:** research paper.

## 1. Introduction

Employee motivation is an integral part of successful business management. In a dynamic business environment where competition is increasingly fierce, keeping employees engaged and highly motivated becomes a key challenge for managers and organisational leaders. A properly motivated staff not only contributes to better performance, but also influences the atmosphere in the workplace, which has a significant impact on organisational culture and company image (Karas, 2004).

In today's professional environment, both financial and intangible aspects play an important role in motivating employees, aiming to increase their commitment and efficiency. The development of the work-life balance concept is becoming increasingly important for employee well-being. The impact of motivators on this balance is extremely important, as it has a direct impact on the quality of employees' professional and personal lives (Lesniewski, Berny, 2011; Kozminski, Piotrkowski, 2007).

Today, both financial and non-financial motivators are an integral part of human resource management strategies. When looking for new jobs, employees not only pay attention to salary, but also to a comprehensive benefits package, which may include flexible working hours, a remote working option, support with health issues and development programmes (Lesniewski, Berny, 2011).

Labour market research shows that organisations that effectively combine a variety of motivators in their offerings enjoy higher employee engagement and loyalty. In this perspective, the concept of work-life balance is becoming increasingly valued. For many employees, it is crucial for maintaining mental and physical health, building family relationships and pursuing passions and interests outside the workplace. Organisations that understand the importance of work-life balance not only promote a healthier and more productive work environment, but also build a positive image as an employer that cares about the well-being of its employees (Paszkiewicz, Wasiluk, 2022; Wiradendi et al., 2021).

Motivators, both financial and non-financial, have a direct impact on achieving and maintaining a work-life balance. By offering employees the opportunity to tailor their work schedules to their individual needs and by making available support programmes or other life-enhancing initiatives, organisations create an environment conducive to harmoniously combining work and private responsibilities. Financial incentives, such as raises or bonuses, can stimulate employees' motivation to perform better. However, when they become the sole or main motivating factor, they can lead to work overload and an imbalance between work and personal life. Employees may be inclined to devote more time and energy to work at the expense of their personal lives, which in the long term may result in job burnout and health problems (Paszkiewicz, Wasiluk, 2022; Kocot, Kwasek, 2023).

On the other hand, non-financial motivators, such as flexible working hours or the possibility to work remotely, can be beneficial in achieving a better work-life balance. They allow employees to adapt their work schedule to their individual needs and life priorities, enabling them to manage their time effectively, engage in non-work activities and take care of their health and well-being. It is important to strike a balance between financial and non-financial motivators in the context of work-life balance. Organisations should strive to create a conducive working environment that takes into account both these spheres of motivation, supporting employees to achieve professional success without negatively impacting on their personal lives. After all, the long-term success of an organisation depends on the health, satisfaction and commitment of its employees, and properly aligned motivators play a key role in this (Tomaszewska-Lipiec, 2014).

## 2. Research material and method

The aim of this study was to analyse the role of employee motivation in effective business management and to identify the main challenges in providing it. It assessed how employees evaluate the various motivational factors and what strategies managers can use to effectively manage the organisation.

A survey method was used in the study. The research tool used was a questionnaire. A proprietary questionnaire consisting of two sections was developed to implement the survey. The first section included socio-demographic data such as age, gender, place of residence and length of service. The second section of the questionnaire concerned the evaluation of the motivation system and the motivators used.

The study assessed the following research problems:
1. How do employees assess their own work motivation?
2. Which motivation instruments do employees most prefer?

The survey was conducted between May and July 2023, using the Internet as the main data collection tool. The survey questionnaire was prepared using the Google Forms platform, and a link to it was made available on various thematic forums and other online platforms. Respondents were assured of the anonymity of their answers and the purpose of the survey. In addition, participants were given the option to stop completing the questionnaire at any time.

The collected results were statistically analysed and presented in the form of tables and graphs. The method of frequency analysis and basic descriptive statistics for quantitative data, such as mean, median and standard deviation, were used to analyse the self-reported data.

An analysis of variance (ANOVA) with repeated measures was performed to compare motivator ratings. Comparisons between two independent samples were made using the Student's t-test or the Mann-Whitney test when assumptions about the normality of the data distribution were not met. On the other hand, for comparison of values between more than two independent groups, analysis of variance (ANOVA) or the Kruskal-Wallis test was used. When significant differences were detected, POST-HOC tests were used for more detailed analysis. The normality of the data distribution was checked using the Shapiro-Wilk test. Analysis of the relationship between quantitative variables was performed using Pearson's or Spearman's correlation analysis. The significance level was taken as $\alpha = 0.05$. All analyses were performed using Statistica 13.3 software from StatSoft.

There were 102 participants in the self-survey, of whom 61.76% were female and 38.24% were male. The age range of 18 to 25 years was 9.80% of respondents, 26 to 35 years was 48.04% and 36 to 45 years was 29.41%. The remaining 12.75% were employees aged 46 and over. Within the surveyed group, the largest age group were respondents with a length of service of 11 to 15 years and 16 to 20 years. Seniority of more than 20 years was indicated less frequently. On the other hand, seniority of 6 to 10 years was declared by 9.80% of the respondents and less than 5 years by 5.88%.

## 3. Research results and discussion

The participants were subjected to a self-assessment of their level of motivation, which showed that half of them assessed their level of motivation as medium (50.00%). In the study group, 39.22% of the subjects described their motivation as high or very high, while the remaining subjects indicated a low (7.84%) or very low level of self-motivation. Statistically significant differences in the level of work motivation were observed between women and men ($p = 0.005$). Work motivation was found to be significantly higher among men ($M = 3.82$; $SD = 0.91$) compared to women ($M = 3.16$; $SD = 0.88$). Detailed results are shown in table 1.

**Table 1**.
*Self-assessment of level of motivation by gender (N = 102)*

| Gender | Woman ($n = 63$) | | | Man (n = 39) | | | Significance |
|---|---|---|---|---|---|---|---|
| Self-assessment of level of motivation | *M* | *Me* | *SD* | *M* | *Me* | *SD* | *p* |
| | 3,16 | 3,00 | 0,88 | 3,82 | 3,00 | 0,91 | 0,005* |

\* p < 0,01; M – Mean; Me – Median; SD – Standard Deviation; p – Probability Level.

Source: Own elaboration based on conducted research.

Statistically significant differences were also shown between age groups in terms of the level of motivation to work ($p = 0.016$). The highest level of motivation was observed among the oldest people ($M = 3.85$; $SD = 1.68$), while the lowest level of motivation was shown in the group of people aged 26-35 ($M = 3.20$; $SD = 0.84$) (Table 2).

**Table 2.**
*Self-assessment of level of motivation by age (N = 102)*

| Self-assessment of level of motivation | M | Me | SD | p |
|---|---|---|---|---|
| 18-25 years (n = 10) | 3,70 | 4,00 | 0,48 | |
| 26-35 years (n = 49) | 3,20 | 3,00 | 0,84 | 0,016** |
| 36-45 years (n = 30) | 3,47 | 3,00 | 0,73 | |
| 46 years and more (n = 13) | 3,85 | 5,00 | 1,68 | |

\*\* p < 0,05; M - mean; Me - median; SD - standard deviation; p - probability level

Source: Own elaboration based on conducted research

The analysis also showed statistically significant differences in the level of motivation according to length of service (p < 0.001). The highest level of motivation was found among employees with more than 20 years of work experience (M = 4.12; SD = 1.54), while the lowest level of motivation was found among those working for less than 5 years (M = 3.00; SD = 0.01) (Table 3).

**Table 3.**
*Self-assessment of level of motivation by seniority (N = 102)*

| Self-assessment of level of motivation | M | Me | SD | p |
|---|---|---|---|---|
| Below 5 years (n = 6) | 3,00 | 3,00 | 0,01 | |
| 6-10 years (n = 10) | 3,70 | 4,00 | 0,48 | |
| 11-15 years (n = 37) | 3,41 | 3,00 | 0,83 | <0,001*** |
| 16-20 years (n = 32) | 3,09 | 3,00 | 0,59 | |
| Above 20 years (n = 17) | 4,12 | 5,00 | 1,54 | |

\*\*\* p < 0,001; M - mean; Me - median; SD - standard deviation; p - probability level.

Source: Own elaboration based on conducted research.

The majority of respondents (82.35% in total) declared that they had received (30.39%) or rather received (51.96%) this information regarding the workplace motivation system. In contrast, a smaller group of respondents (17.65% in total) stated that they rather did not have such knowledge (6.86%) and definitely did not know the motivation system used at the workplace (10.78%).

The respondents rated the importance of financial motivators on a scale of 1 to 5, where 1 meant low importance and 5 meant very high importance. The results obtained are presented in Table 4. The analysis showed statistically significant differences in the rating of individual motivators (p = 0.013). Financial motivators were ranked from lowest to highest rated. The highest rated financial motivator was the awarding of bonuses for achieving goals (M = 4.60; SD = 0.69), while the lowest rated was receiving rewards for achieving goals (M = 4.19; SD = 1.21).

**Table 4.**
*Evaluation of the effectiveness of financial motivators (N = 102)*

| Financial motivator | M | Me | Min | Max | SD | F | p |
|---|---|---|---|---|---|---|---|
| Rewards for achieving targets [a] | 4,19 | 5,00 | 1,00 | 5,00 | 1,21 | 3,91 | 0,013 ** |
| Percentage of sales [ab] | 4,23 | 5,00 | 1,00 | 5,00 | 1,19 | | |
| Salary [ab] | 4,41 | 5,00 | 2,00 | 5,00 | 0,73 | | |
| Bonus for meeting targets [b] | 4,60 | 5,00 | 3,00 | 5,00 | 0,69 | | |

[abc] successive letters stand for homogeneous groups; ** $p < 0,05$; M – mean; Me – median; SD – standard deviation; Min – minimum value; Max – maximum value; F – test statistic; p – probability level.

Source: Own elaboration based on conducted research.

Respondents were asked to rate the effectiveness of non-financial motivators on a scale of 1 to 5, where 1 meant low effectiveness and 5 meant very high effectiveness. The results obtained are presented in Table 5. The analysis showed statistically significant differences in the rating of individual motivators ($p < 0.001$). It was found that stocks and bonds were considered the least effective (M = 2.02; SD = 1.39). In contrast, extra days off (M = 4.10; SD = 1.27), subsidised holidays (M = 4.10; SD = 1.32) and the possibility to use a company car, fuel or reimbursement of commuting costs (M = 4.18; SD = 0.99) were considered to be the most effective.

In addition, statistically significant differences were found in the evaluation of individual intangible motivators ($p < 0.001$). Public praise was rated lowest (M = 2.98; SD = 1.36). In contrast, flexible work schedules (M = 4.29; SD = 0.85), opportunities for career development and advancement (M = 4.37; SD = 0.86) and work atmosphere and comfort (M = 4.46; SD = 0.78) were rated highest.

**Table 5.**
*Evaluation of the effectiveness of non-financial motivators (N = 102)*

| Non-financial motivator | M | Me | Min | Max | SD | F | p |
|---|---|---|---|---|---|---|---|
| Shares, bonds [a] | 2,02 | 1,00 | 1,00 | 5,00 | 1,39 | | |
| Company housing [ab] | 2,46 | 2,00 | 1,00 | 5,00 | 1,65 | | |
| Staff loans [abc] | 2,68 | 2,00 | 1,00 | 5,00 | 1,57 | | |
| Special events [abc] | 2,78 | 3,00 | 1,00 | 5,00 | 1,38 | | |
| Gym passes, theatres, cinemas, swimming pools [abcd] | 3,07 | 3,00 | 1,00 | 5,00 | 1,48 | | |
| Reimbursement for studies, courses [bcd] | 3,09 | 3,00 | 1,00 | 5,00 | 1,62 | | |
| Company computer, telephone [bcde] | 3,23 | 3,50 | 1,00 | 5,00 | 1,55 | 9,23 | 0,001*** |
| Lunches, buffet, catering [bcde] | 3,29 | 4,00 | 1,00 | 5,00 | 1,59 | | |
| Supplementary insurance (group, life) [bcde] | 3,42 | 3,00 | 1,00 | 5,00 | 1,15 | | |
| Medical care [cde] | 3,86 | 4,00 | 1,00 | 5,00 | 1,02 | | |
| Christmas parcels or vouchers [de] | 3,90 | 4,00 | 1,00 | 5,00 | 1,26 | | |
| Additional days off [e] | 4,10 | 5,00 | 1,00 | 5,00 | 1,27 | | |
| Holiday allowance [e] | 4,10 | 5,00 | 1,00 | 5,00 | 1,32 | | |
| Company car, fuel or reimbursement of commuting expenses [e] | 4,18 | 4,00 | 1,00 | 5,00 | 0,99 | | |

[abc] successive letters stand for homogeneous groups; *** $p < 0,001$; M – mean; Me – median; SD – standard deviation; Min – minimum value; Max – maximum value; F – test statistic; p – probability level.

Source: Own elaboration based on conducted research.

**Table 6.**
*Evaluation of the effectiveness of intangible motivators (N = 102)*

| Intangible motivator | M | Me | Min | Max | SD | F | p |
|---|---|---|---|---|---|---|---|
| Public praise [ab] | 2,98 | 3,00 | 1,00 | 5,00 | 1,36 | 15,22 | <0,001*** |
| Training and coaching [bc] | 3,67 | 4,00 | 1,00 | 5,00 | 1,32 | | |
| Work-life balance [cd] | 4,15 | 5,00 | 1,00 | 5,00 | 1,16 | | |
| Professional development [cd] | 4,24 | 5,00 | 2,00 | 5,00 | 0,96 | | |
| Flexible schedule [d] | 4,29 | 5,00 | 3,00 | 5,00 | 0,85 | | |
| Opportunities for development and promotion [d] | 4,37 | 5,00 | 2,00 | 5,00 | 0,86 | | |
| Atmosphere and comfort at work [d] | 4,46 | 5,00 | 3,00 | 5,00 | 0,78 | | |

[abc] successive letters stand for homogeneous groups; *** p < 0,001; M – mean; Me – median; SD – standard deviation; Min – minimum value; Max – maximum value; F – test statistic; p – probability level.

Source: Own elaboration based on conducted research.

## 4. Summary

In today's dynamic labour market, which is saturated with competition, there is the challenge of maintaining a work-life balance. Organisations that are aware of these challenges and take initiatives to promote this balance can reap numerous benefits, including increased employee motivation. Promoting work-life balance signals an organisation's concern for the wellbeing of its employees, which builds trust and loyalty among staff. As a result, these employees are often more committed to their responsibilities, loyal to the company and willing to engage in additional activities that contribute to the success of the organisation (Knap-Stefaniuk, 2018; Mroczkowska, Kubacka, 2020).

The results of the present study indicate a diversity of motivation levels among employees, which is an important aspect in the context of human resource management and the development of effective motivational strategies. A surprisingly high proportion of employees (50%) assess their motivation as average. The results obtained in this study confirm that there is great potential to improve motivation in many organisations.

The study showed differences in motivation by gender, age and length of service. Men often show higher levels of motivation than women, which may be due to differences in career expectations or availability of resources. Older people tend to have higher levels of motivation, which may be related to work experience and a sense of professional fulfilment. In contrast, younger employees may experience lower levels of motivation due to lack of development prospects or work-life conflict (Paszkiewicz, Wasiluk, 2022).

The final aspect of seniority also highlights the relevance of work experience in terms of motivation. Those working for more than 20 years can derive satisfaction from long-term commitment and visible contribution to the organisation. In contrast, younger employees, working less than five years, may experience initial adaptation difficulties or a lack of understanding of their needs and expectations.

Financial factors, such as remuneration and financial rewards, are considered the most important factors in motivating employees. Non-financial factors, in particular job security, also play an important role in motivating employees. The analysis of variance suggests that a combination of financial and non-financial motivators is necessary to achieve the desired effects in employee motivation, which is also supported by academic research (Rakić et al., 2022).

The results of the survey clearly indicate the dominant role of the target achievement bonus as the main financial motivator among respondents. Modern organisations are increasingly using a variety of incentive systems to increase employee engagement and productivity, with financial motivators playing an important role in this context. A high rating of bonuses for the achievement of specific goals may be indicative of several key aspects. Firstly, it may suggest that employees are aware of the demands of the organisation and value clearly defined standards and expectations. Bonuses, as a form of direct reward for performance, may be perceived by employees as fair and appropriate to the effort put in. In addition, effective bonus systems can stimulate competitiveness in the workplace and encourage employees to strive for continuous development and improvement of their skills. However, despite the dominance of financial motivators, it is also important to consider other motivational aspects, such as professional development, job satisfaction or a positive team atmosphere, which can have an equally significant impact on employee engagement and satisfaction (Rakić et al., 2022).

The results of the presented study clearly show that non-financial motivators are an important element in shaping employee satisfaction and commitment. The conclusion from our research is the low effectiveness of shares and bonds as motivational tools. Traditionally, it was thought that participation in company profits or the opportunity to invest in the company should be an attractive motivator. However, the current results suggest that, in light of alternative forms of compensation, this approach may no longer be so encouraging for employees. It may also indicate employees' concerns about the stability and future of the company. On the other hand, the high effectiveness of additional days off, holiday subsidies or the possibility to use a company car and reimburse commuting expenses highlights the importance of work-life balance. The contemporary labour market increasingly emphasises flexibility and customisation for employees, which is supported by our results showing the high price of such benefits (Wiradendi et al., 2021).

Our study on the evaluation of intangible motivators revealed several important findings regarding the value employees place on different aspects of their work. Surprisingly, public praise, which is often considered a simple and inexpensive way to increase employee motivation, was rated as the least effective. Such a result may indicate that in today's work environment, employees value individual recognition and feedback more, which is not presented publicly. Public forms of praise may be perceived as less authentic or even stressful for some employees, which affects their perception of work quality. In addition, high ratings for flexible schedules, opportunities for development and promotion, and work atmosphere and comfort emphasise the importance of work-life balance and long-term professional development. Today's employees are increasingly looking for workplaces that offer them not only tangible financial benefits, but also opportunities for development, flexibility in working hours and a friendly, supportive atmosphere (Menderak, 2019).

In conclusion, in order to effectively motivate employees, organisations should consider both financial and non-financial factors and tailor motivational strategies to individual staff needs and expectations.